\documentclass{ws-ijmpa}
\usepackage[super,compress]{cite}
\usepackage{epsfig}
\usepackage{dcolumn}
\usepackage{bm}
\usepackage{amsmath}
\usepackage{amsfonts}
\usepackage{amssymb}
\usepackage{graphicx}
\usepackage{color}
\usepackage{latexsym}
\usepackage{slashed} 

\begin{document}

%
\catchline{}{}{}{}{}
%
\title{SUPERSYMMETRY BREAKING IN ANTI-DE SITTER SPACETIME}

\author{Bin Zhu}

\address{School of Physics, Nankai University, Tianjin 300071,
China\\
binzhu7@gmail.com}

\author{Kun Meng}

\address{School of Science, Tianjin Polytechnic University, Tianjin 300071,
China\\
kunmeng@mail.nankai.edu.cn}

\author{Ran Ding}

\address{School of Physics, Nankai University, Tianjin 300071,
China\\
dingran@mail.nankai.edu.cn}

\maketitle

\begin{history}
\received{Day Month Year}
\revised{Day Month Year}
\end{history}

\begin{abstract}
We study the questions of how supersymmetry is spontaneously broken in Anti de-Sitter spacetime. We verify that the would-be R-symmetry in $AdS_4$ plays a central role for the existence of meta-stable supersymmetry breaking. To illustrate, some well-known models such as Poloyni models and O'Raifeartaigh models are investigated in detail. Our calculations are reliable in flat spacetime limit and confirm us that meta-stable vacua are generic even though quantum corrections are taken into account.


\end{abstract}
\ccode{PACS numbers: 11.10.-z, 11.30.Pb, 11.30.Qc }
\section{Introduction}

Supersymmetry and its dynamical breaking in flat spacetime have been extensively investigated, for reviews, see Refs. \refcite{Shadmi:1999jy,Intriligator:2007zz,Dine:2010cv}. Their characters such as calculability and metastability can be unambiguously computed in a consistent and simple framework of O'Raifearitaigh model \refcite{Intriligator:2006dd,Intriligator:2007py,Giveon:2007ef,Dine:2007dz,Giveon:2008ne,Kitano:2008tm}, when we accept the metastable vacua from the beginning. Dedicated analysis on these models, including computing Coleman-Weinberg potential for pseudomoduli, taught us a great deal on the subtleties of the model building, especially on the lifetime of meta-stable vacuum. As an example, ISS model, the phenomenologically long-lived lifetime of meta-stable vacuum is guaranteed by the dimensionless parameter $\epsilon$
\begin{align}
\epsilon=\frac{\mu}{\Lambda_m} ,
\end{align}
which is the ratio of mass of quarks and dynamical scale. In this work we examine a similar problem in Anti de-Sitter spacetime ($AdS_4$), i.e., SUSY breaking and its metastability in deformed background. This issue has been investigated in Ref.\refcite{McArthur:2013wv} based on Goldstino effective action\cite{Komargodski:2009rz}. Here we concentrate directly on the model construction of SUSY breaking and demonstrate manifestly that meta-stable vacuum is generic without R-symmetry.

In the following part of this section, we present the basic methodology of this paper and figure out the physical significance of it. The studies of supersymmetry in $AdS_4$ were first carried out in \refcite{Keck}, then generalized to Wess-Zumino model in \refcite{Ivanov} and conformal field theory in \refcite{Aharony:2010ay}. Exhaustive studies on this subject \refcite{Adams:2011vw,Festuccia:2011ws,Kapustin:2011gh,Kapustin:2011vz,Kapustin:2012iw} show explicitly that conditions for preserving supersymmetry are
\begin{align}
W_i+\lambda K_i=0,
\label{F-flatness}
\end{align}
and corresponding scalar potential is
\begin{align}
V=g^{i\bar j}(\partial_i W+\lambda \partial_i K)(\partial_{\bar j} \bar W+\lambda \partial_{\bar j} K)-3\lambda W-3\lambda\bar W-3\lambda^2K.
\end{align}
Eq.(\ref{F-flatness}) constitute $n$ equations for $n$ complex variables and generally result in a discrete set of vacua. In addition, the transformation law for chiral multiplets in $AdS_4$ is listed as follows
\begin{align}
\delta_{\xi}\phi^{i}&=\sqrt{2}\xi\psi^i,\\
\delta_{\xi}\chi^{i}&=\sqrt{2}F^{i}\xi+i\sqrt{2}\sigma^{m}\bar\xi\partial_{m}\phi^i,\\
\delta_{\xi}F^{i}&=-\sqrt{2}\lambda\xi\chi^i+i\sqrt{2}\bar\xi\bar\sigma^{m}\nabla_{m}\chi^{i}.
\end{align}
In order that supersymmetry algebra in $AdS_4$ closes on these fields, killing spinor equation for $\xi$ must be imposed
\begin{align}
(\nabla_m\xi)^{\alpha}=\frac{i\lambda}{2}
(\bar\xi\bar\sigma_m)^{\alpha}.
\label{killing spinor}
\end{align}
Since there is no chiral rotation for $\xi$ which can be seen from eq.(\ref{killing spinor}). That means R-symmetry is explicitly broken in $AdS_4$. Recall that R symmetries play a crucial role for SUSY breaking collected into a theorem of Nelson and Seiberg\cite{Nelson:1993nf}: In order that a generic lagrangian breaks supersymmetry, the theory must possess an R symmetry. Here we demonstrate manifestly that the theorem still holds in $AdS_4$ in terms of some well-known models.

In section $2$, we set up a general analysis on Poloyni model and its interesting deformation. We observe that SUSY breaking with single chiral superfield is impossible unless $m=-\lambda$ is permitted. Meta-stable SUSY breaking is still available when we take a special hierarchy between couplings.

In section $3$, we focus our attention on the simplest O'Raifeartaigh model with superpotential
\begin{align}
W=fX+\frac{1}{2}hX\phi_1^2+m\phi_1\phi_2.
\end{align}
We find that in the flat spacetime limit, quantum corrections generally result in local minima near the origin. Interestingly, this model is identical to the deformation of Poloyni model when we match the parameters appearing in the model.

In section $4$, we construct models with non-canonical Kahler potential. We show that for a wide range of parameters, there are no supersymmetric vacua and thus spontaneously break supersymmetry. Finally, we provide some general comments on ISS model when it is put in $AdS_4$.

\section{Poloyni Model in $AdS_4$}
The considerations of SUSY breaking in flat spacetime are illustrated by Poloyni model with linear superpotential
\begin{align}
K=X\bar X, \quad W=f X.
\end{align}
An important point in implementing Poloyni model for $AdS_4$ should be noted. We mentioned that decomposition to real and imaginary parts of chiral superfield is favored for dealing with complicated theories.
\begin{align}
X=\frac{1}{\sqrt{2}}\left(a+ib\right).
\end{align}
Supersymmetry is restored obviously from deformed F-flatness conditions (\ref{F-flatness}),
\begin{align}
a= - \frac{\sqrt{2}f}{\lambda },\quad b= 0.
\end{align}
Moving to general Wess-Zumino model, we follow the computation procedure of Poloyni model. We found that SUSY breaking model is highly constrained in $AdS_4$, i.e., there is only one model violating (\ref{F-flatness}).
\begin{align}
K=X\bar X,\quad W=fX+\frac{1}{2}mX^2+\frac{1}{3}hX^3.
\end{align}
After detailed computation, we have
\begin{align}
a&=\frac{\lambda -m}{\sqrt{2} h},\quad b=
   -\frac{i \sqrt{-4 f h-3 \lambda
   ^2+m^2+2 \lambda  m}}{\sqrt{2}h};\\
a&=\frac{\lambda-m}{\sqrt{2} h},\quad b= \frac{i \sqrt{-4 f
   h-3 \lambda ^2+m^2+2 \lambda
   m}}{\sqrt{2}h};\\
a&=\frac{-\sqrt{-4 f h+ \lambda ^2+
   m^2+2 \lambda  m}- \lambda
   -m}{\sqrt{2} h},\quad b=
   0;\\
a&=\frac{\sqrt{-4 f h+ \lambda ^2+
   m^2+2 \lambda  m}- \lambda
   -m}{\sqrt{2} h},\quad b= 0.
\end{align}
Generically, there are always supersymmetric vacua due to the explicitly broken R-symmetry. Supersymmetry can be broken if the following constraints are satisfied
\begin{align}
-4 f h + m^2 + 2 m \lambda - 3 \lambda^2 >
  0,\quad -4 f h +  m^2 + 2 m \lambda +  \lambda^2 < 0.
\label{constraints}
\end{align}
 Obviously, these two equations are not compatible. SUSY breaking appears to be impossible for Wess-Zumino model. The novelty here is that once we set the coupling $h$ to be zero from the outset. Subsequent study shows that it would induce supersymmetry breaking by adapting the choice of parameter:
 \begin{align}
 m&=-\lambda,\\
K&=X\bar X,\quad W=fX-\frac{1}{2}\lambda X^2.
\end{align}
 We see that in this type of superpotential, the complex scalar becomes massless $m^2=0$, which is above the Breitenlohner-Freedman bound $-9/4\lambda^2$\cite{Breitenlohner:1982jf}. The spontaneously broken SUSY vacuum is thus stable in $AdS_4$. It is easy to generalize to two-field situation
\begin{align}
K=\phi_1\bar\phi_1+\phi_2\bar\phi_2,\quad W=\lambda  \phi _1 (f+\phi
   _2).
\end{align}
So far we have examined global properties of SUSY breaking in $AdS_4$, let us start to analyze the local version of meta-stable SUSY breaking. The scalar potential written in terms of $a$ and $b$ for Wess-Zumino model
\begin{align}
V=&\frac{a^4 h^2}{4}+\frac{a^3 h
   m}{\sqrt{2}}+\frac{1}{2} a^2 b^2
   h^2+a^2 f h-a^2 \lambda
   ^2+\frac{a^2 m^2}{2}-\frac{1}{2}
   a^2 \lambda  m+\frac{a b^2 h
   m}{\sqrt{2}}\nonumber\\&-2 \sqrt{2} a f
   \lambda +\sqrt{2} a f m+\frac{b^4
   h^2}{4}-b^2 f h-b^2 \lambda
   ^2+\frac{b^2 m^2}{2}+\frac{1}{2}
   b^2 \lambda  m+f^2.
\end{align}
The solutions satisfying $\partial V=0$ are
\begin{align}
&a= \frac{
   \lambda - m}{\sqrt{2}h},\quad b=
   -\frac{i \sqrt{-4 f h-3 \lambda
   ^2+m^2+2 \lambda  m}}{\sqrt{2}
   h};\nonumber\\
&a=
   \frac{\lambda -
   m}{\sqrt{2} h},\quad b= \frac{i \sqrt{-4 f
   h-3 \lambda ^2+m^2+2 \lambda
   m}}{\sqrt{2}
   h};\nonumber\\
&a= \frac{2
   \lambda -m}{\sqrt{2}
   h},\quad b= 0;\\
a&=\frac{-\sqrt{-4 f h+ \lambda ^2+
   m^2+2 \lambda  m}- \lambda
   -m}{\sqrt{2} h},\quad b=
   0;\nonumber\\
a&=\frac{\sqrt{-4 f h+ \lambda ^2+
   m^2+2 \lambda  m}- \lambda
   -m}{\sqrt{2} h},\quad b= 0.\nonumber
\end{align}
There is one additional solution which is not SUSY vacua
\begin{align}
a=\frac{2 \lambda
   -m}{\sqrt{2} h},\quad b=0.
\end{align}
 According to the multi-value function stationary point theorem, we learn that if $z=f(x,y)$ is continuous at $z_0$, and exist second-derivative, $f_x(x_0,y_0)=f_y(x_0,y_0)=0$. Denote
\begin{align}
A=f_{xx}(x_0,y_0),\quad B=f_{xy}(x_0,y_0),\quad C=f_{yy}(x_0,y_0),
\end{align}
\begin{enumerate}
\item If $AC-B^2>0$ and $A>0$, it is minima,
\item If $AC-B^2>0$ and $A<0$, it is maxima.
 \end{enumerate}
For Wess-Zumino model
\begin{align}
A&=2 f h-\frac{1}{2} (m-2 \lambda ) (4
   \lambda +m)>0,\nonumber\\
AC-B^2&=\left(\frac{1}{2} m (2 \lambda +m)-2
   f h\right) \left(2 f h-\frac{1}{2}
   (m-2 \lambda ) (4 \lambda
   +m)\right)>0.
   \label{Constraint1}
\end{align}
Additionally, we require the tunneling amplitude
\begin{align}
\Gamma\sim\frac{\Delta V}{\Delta\phi}=\frac{\left((2 \lambda +m)^2-4 f
   h\right) \left(\sqrt{(\lambda
   +m)^2-4 f h}+5 \lambda \right)}{8
   \sqrt{2} h}-\frac{3 \lambda
   ^3}{\sqrt{2} h},
   \label{Constraint2}
\end{align}
is highly suppressed. Combine the three constraints (\ref{Constraint1})(\ref{Constraint2}) over the parameters of the model, we find there is actually a parameter range satisfying the above constraints
\begin{align}
m\sim2\sqrt{fh},\quad\frac{\sqrt{fh}}{2}<\lambda<\frac{4\sqrt{fh}}{3},
\end{align}
with $f$ and $h$ being small. Upon to this point, we mentioned that meta-stable SUSY breaking is present.

Consider the behavior of Poloyni model with deformed Kahler potential
\begin{align}
K=X\bar X-\frac{1}{\Lambda^2}(X\bar X)^2,\quad W=fX.
\label{Metastable}
\end{align}
This theory is not renormalizable. It should be regarded as quantum field theory with cutoff $\Lambda$. At first glance, the conditions for supersymmetric vacua become
\begin{align}
-a^3-a b^2 +a
   \Lambda^2+\frac{\sqrt{2}f\Lambda^2}{\lambda}=0,\quad b(a^2
   +b^2-\Lambda ^2)
   =0.
   \label{F}
\end{align}
Taking a special hierarchy suppresses the tunneling amplitude
\begin{align}
\lambda <<\sqrt{f}<\Lambda.
\end{align}
And redefine
\begin{align}
\lambda&=\sqrt{f} \epsilon
   _1,\quad
\Lambda=\frac{\sqrt{f}}{\epsilon
   _2},
\label{hierarchy}
\end{align}
where underlying assumption is  $\epsilon_1<<\epsilon_2$. Substituting into (\ref{F}) simplifies the derivation of supersymmetric vacua.
\begin{align}
a\sim
   \frac{\sqrt{f}}{\epsilon
   _1^{1/3} \epsilon _2^{2/3}},\quad b=0.
\end{align}
Expand the scalar potential to the order $O(\epsilon_1)$ and $O(\epsilon_2^2)$ respectively
\begin{align}
V=2f\epsilon _2^2 \left(a^2+ b^2
   \right)-2 \sqrt{2} a f^{3/2}
   \epsilon _1+f^2.
   \label{SimplifiedV}
\end{align}
Differentiate eq.(\ref{SimplifiedV}) with a and b, we get the stationary points of scalar potential near the origin
\begin{align}
a\sim \sqrt{f}\left(\frac{ \epsilon
   _1}{\epsilon _2}\right)\frac{1}{\epsilon _2}.
\end{align}
When the hierarchy (\ref{hierarchy}) is taken, the analysis is simplified and framed in a conventional language that meta-stable vacuum is stabilized due to the fact that it is far away from supersymmetric vacua.

\section{O'Raifeartaigh Model in $AdS_4$}
Our main example of meta-stable supersymmetry breaking in this paper is greatly simple: O'Raifeartaigh model
\begin{align}
K=X\bar X+\phi_1\bar\phi_1+\phi_2\bar\phi_2,\quad W=fX+\frac{1}{2}hX\phi_1^2+m\phi_1\phi_2.
\end{align}
Supersymmetry is restored as Poloyni model in $AdS_4$. In order to have control over our theory, we take all parameters appearing in theory greatly larger than $\lambda$. In this scenario, theory is still treated as flat spacetime Supersymmetry, only the vacuum is deformed. Therefore, quantum correction is the same as flat spacetime. We can regard it as either Coleman-Weinberg potential\cite{Coleman:1973jx} or Effective Kahler potential\cite{Grisaru:1996ve,Benjamin:2010vm,Flauger:2012ie}. For simplicity, we choose the  second one. The general procedure is listed as follows,
\begin{enumerate}
\item Find pseudomoduli in the original superpotential without considering curvature effects. Here $X$ is pseudomoduli.
\item Fix the field value to derive the reduced superpotential.
The original O'Raifeartaigh model is then reduced to $W=fX$
\item Quantum corrections are collected into the effective Kahler potential.
    \begin{align}
    K_{eff}=K_{tree}+K_{1-loop}\;,\quad K_{1-loop}=-\frac{1}{32\pi^2}Tr\left[\mathcal{M}\mathcal{M}^{+}\ln
    \frac{\mathcal{M}\mathcal{M}^+}{\Lambda^2}\right].
    \end{align}
    The good point of this method is that it respects the idea of non-renormalization theorem for superpotential. While the bad one is that it is only applicable for the case where SUSY breaking is small
\item Expand the one-loop Kahler potential
\begin{align}
K_{1-loop}=k_{0,0}+k_{2,1}X\bar X+k_{4,2}(X\bar X)^2+\ldots\;,
\end{align}
Notice that $k_{0,0}$ has nothing to do with physics if we decouple dynamical gravity, so it can be ignored safely. $k_{2,1}$ works as correction to inverse Kahler metric, and suppressed by loop-factor, it is also safely ignored. So the relevant effective Kahler potential becomes
\begin{align}
K_{eff}=|X|^2-\frac{1}{32 \pi ^2}\frac{h^3}{6m^2}|X|^4
  +\ldots\;,
  \label{Effective}
\end{align}
\end{enumerate}
 Eq.(\ref{Effective}) is the same as meta-stable SUSY breaking case at eq.(\ref{Metastable}) if we identify
\begin{align}
\frac{1}{\Lambda ^2}=
   \frac{1}{32 \pi ^2}\left(\frac{h^3}{6m^2}\right).
\end{align}
In the simplified description of theory, meta-stable SUSY breaking is induced by including the considerations of quantum corrections.

\section{Non-canonical Kahler Potential}
Working in canonical Kahler potential means $AdS_4$ superalgebra can reduce to Poincare superalgebra at the limit of $\lambda\rightarrow 0$. Abandoning this requirement has very important impact on the distribution of supersymmetric vacua\cite{Sun:2011aq}. The simplest example is
\begin{align}
K=\log(X+\bar X)+X\bar X,\quad W=fX.
\end{align}
Following the same computational procedure, we derive a set of discrete of supersymmetric vacua
\begin{align}
a&=
   \frac{-\sqrt{f^2-2 \lambda
   ^2}-f}
   {\sqrt{2}\lambda },\quad b=
   0;\nonumber\\
a&=
    \frac{\sqrt{f^2-2 \lambda
   ^2}-f}
   {\sqrt{2}\lambda },\quad b=
   0.
\end{align}
Assuming $f^2<2\lambda^2$ ruins the existence of physically supersymmetric vacua. Another example is intended to the deformed O'Raifeartaigh model
\begin{align}
K=\log(X+\bar X)+\phi_1\bar\phi_1+\phi_2\bar\phi_2,\quad W=\frac{1}{2}hX\phi_1^2+m\phi_1\phi_2.
\end{align}
If the mass of fields are larger than $\lambda$, theory will break SUSY without requiring theory has flat spacetime limit.

\section{Comments on ISS model in $AdS_4$}
Recently, Aharony et al. have explored supersymmetric gauge theories in $AdS_4$ with $N=4$\cite{Aharony:2010ay} and $N=1$\cite{Aharony:2012jf} . In flat spacetime, Meson and Baryon are good candidates of degrees of freedom at low energy. When $N_c+1<N_f<\frac{3}{2}N_c$, Seiberg duality\cite{Seiberg:1994pq} guarantees the existence of weakly coupled magnetic theory at IR. In that regime, ISS model has rank-condition SUSY breaking
\begin{align}
f\delta_{ij}+\phi_i\tilde\phi_j.
\end{align}
Moving far away from the origin, non-perturbative effects\cite{Seiberg:1994bp} restore supersymmetry. The dangerous pseudomoduli space is stabilized by Coleman-Weinberg potential. We encounter lots of questions when putting ISS model in $AdS_4$. The first one is that we are not sure whether or not Seiberg duality holds in this background. To define the consistent model in $AdS_4$, boundary conditions should to be imposed. With different types of boundary conditions, coming different physics. For now, we just take a schematic method to analyze the theory and comment the possible results in the flat spacetime limit. There are lots of changes for the moduli space even in this limit. For example, the runaway for $N_f<N_c$ is regulated in $AdS_4$. For ISS model, the rank condition breaking mechanism does not exist,
\begin{align}
f\delta_{ij}+\phi_i\tilde\phi_j+\lambda\Phi^+,
\end{align}
Supersymmetry is not broken in IR. There is no need to include non-perturbative effects. The interesting point is that there is still meta-stable vacua when non-calculable quantum corrections are introduced for Kahler potential as what we have done in O'Raifeartaigh model
\begin{align}
K=Tr\Phi\bar\Phi-\frac{1}{\Lambda_m^2}(Tr\Phi\bar\Phi)^2+\ldots\;,\quad W=f\Phi+\ldots
\end{align}
Based on former experience, we can show that there will be meta-stable vacuum far away from the supersymmetric vacuum.

\section{Conclusion}
Exploring properties of SUSY breaking is interesting topic even though the background is not flat. In this work we show that explicitly broken R-symmetry can restore supersymmetry . We have executed the analysis in various types of models. First we examined the simplest Poloyni model which does not break supersymmetry in $AdS_4$. After doing concrete calculations for Wess-Zumino model, we find supersymmetry can not be broken unless $h=0, m=-\lambda$.  Another way to break supersymmetry is meta-stable SUSY breaking which is achieved by an effective field theory description with cutoff $\Lambda$ and Poloyni-type superpotential. Its UV completion can be regarded as O'Raifeartaigh model in special flat spacetime limit. The interesting point is that some information about SUSY breaking for ISS model in $AdS_4$ can also be obtained in this effective field theory description. Furthermore, if we give up the canonical Kahler potential requirment, SUSY can be broken in global vacuum easily.


\begin{thebibliography}{0}    

\bibitem{Shadmi:1999jy}
  Y.~Shadmi, Y.~Shirman and ,
  Rev.\ Mod.\ Phys.\  {\bf 72} (2000) 25
  [hep-th/9907225].

\bibitem{Intriligator:2007zz}
  K.~A.~Intriligator and N.~Seiberg,
  Class.\ Quant.\ Grav.\  {\bf 24} (2007) S741
  [hep-ph/0702069].

\bibitem{Dine:2010cv}
  M.~Dine and J.~D.~Mason,
  Rept.\ Prog.\ Phys.\  {\bf 74} (2011) 056201
  [arXiv:1012.2836 [hep-th]].

\bibitem{Intriligator:2006dd}
  K.~A.~Intriligator, N.~Seiberg, D.~Shih and ,
  JHEP {\bf 0604} (2006) 021
  [hep-th/0602239].

\bibitem{Intriligator:2007py}
  K.~A.~Intriligator, N.~Seiberg, D.~Shih and ,
  JHEP {\bf 0707} (2007) 017
  [hep-th/0703281].


\bibitem{Giveon:2007ef}
  A.~Giveon, D.~Kutasov and ,
  Nucl.\ Phys.\ B {\bf 796} (2008) 25
  [arXiv:0710.0894 [hep-th]].

\bibitem{Dine:2007dz}
  M.~Dine, J.~D.~Mason and ,
  Phys.\ Rev.\ D {\bf 78} (2008) 055013
  [arXiv:0712.1355 [hep-ph]].


\bibitem{Giveon:2008ne}
  A.~Giveon, A.~Katz, Z.~Komargodski, D.~Shih and ,
  JHEP {\bf 0810} (2008) 092
  [arXiv:0808.2901 [hep-th]].

\bibitem{Kitano:2008tm}
  R.~Kitano, Y.~Ookouchi and ,
  Phys.\ Lett.\ B {\bf 675} (2009) 80
  [arXiv:0812.0543 [hep-ph]].


\bibitem{Keck}
B.~W.~Keck, ``An Alternative Class of Supersymmetries,`` J. Phys. A 8 (1975) 1819

\bibitem{Ivanov}
E. A. Ivanov and A. S. Sorin, ``Superfield Formulation Of Osp(1,4) Supersymmetry`` J. Phys. A 13 (1980) 1159

\bibitem{McArthur:2013wv}
  I.~N.~McArthur,
  arXiv:1301.4842 [hep-th].

\bibitem{Komargodski:2009rz}
  Z.~Komargodski and N.~Seiberg,
  JHEP {\bf 0909} (2009) 066
  [arXiv:0907.2441 [hep-th]]


\bibitem{Adams:2011vw}
  A.~Adams, H.~Jockers, V.~Kumar and J.~M.~Lapan,
  JHEP {\bf 1112} (2011) 042
  [arXiv:1104.3155 [hep-th]].

\bibitem{Festuccia:2011ws}
  G.~Festuccia and N.~Seiberg,
  JHEP {\bf 1106} (2011) 114
  [arXiv:1105.0689 [hep-th]].

\bibitem{Kapustin:2011gh}
  A.~Kapustin,
  arXiv:1104.0466 [hep-th].

\bibitem{Kapustin:2011vz}
  A.~Kapustin, H.~Kim and J.~Park,
  JHEP {\bf 1112} (2011) 087
  [arXiv:1110.2547 [hep-th]].

\bibitem{Kapustin:2012iw}
  A.~Kapustin, B.~Willett and I.~Yaakov,
  arXiv:1211.2861 [hep-th].

\bibitem{Nelson:1993nf}
  A.~E.~Nelson and N.~Seiberg,
  Nucl.\ Phys.\ B {\bf 416} (1994) 46
  [hep-ph/9309299].

\bibitem{Breitenlohner:1982jf}
  P.~Breitenlohner, D.~Z.~Freedman and ,
  Annals Phys.\  {\bf 144} (1982) 249.

\bibitem{Coleman:1973jx}
  S.~R.~Coleman and E.~J.~Weinberg,
  Phys.\ Rev.\ D {\bf 7} (1973) 1888.

\bibitem{Grisaru:1996ve}
  M.~T.~Grisaru, M.~Rocek and R.~von Unge,
  Phys.\ Lett.\ B {\bf 383} (1996) 415
  [hep-th/9605149].

\bibitem{Benjamin:2010vm}
  S.~Benjamin, C.~Freund and B.~Kain,
  Nucl.\ Phys.\ B {\bf 842} (2011) 529
  [arXiv:1003.5628 [hep-ph]].

\bibitem{Flauger:2012ie}
  R.~Flauger, S.~Hellerman, C.~Schmidt-Colinet and M.~Sudano,
  arXiv:1205.3492 [hep-th].

\bibitem{Sun:2011aq}
  Z.~Sun,
  JHEP {\bf 1109} (2011) 107
  [arXiv:1105.3172 [hep-th]].

\bibitem{Aharony:2010ay}
  O.~Aharony, D.~Marolf and M.~Rangamani,
  JHEP {\bf 1102} (2011) 041
  [arXiv:1011.6144 [hep-th]].
\bibitem{Aharony:2012jf}
  O.~Aharony, M.~Berkooz, D.~Tong and S.~Yankielowicz,
  arXiv:1210.5195 [hep-th]

\bibitem{Seiberg:1994pq}
  N.~Seiberg,
  Nucl.\ Phys.\ B {\bf 435}, 129 (1995)
  [hep-th/9411149].

\bibitem{Seiberg:1994bp}
  N.~Seiberg,
  hep-th/9408013.

\bibitem{Intriligator:2007cp}
  K.~A.~Intriligator and N.~Seiberg,
  Class.\ Quant.\ Grav.\  {\bf 24}, S741 (2007)
  [hep-ph/0702069].
\end{thebibliography}
\end{document}